\documentclass[11pt]{article}

\title{Comments on arXiv:1811.00154 [astro-ph.IM] \\ ``AGN Variability Analysis Handbook''}
\author{Roberto Vio \\ Chip Computers Consulting s.r.l. \\ Viale Don L.~Sturzo 82,
              S.Liberale di Marcon, 30020 Venice, Italy \\ \\
Paola Andreani \\ ESO \\ Karl Schwarzschild strasse 2, 85748 Garching, Germany\\ \\}

\begin{document}
\maketitle

\abstract{{\bf Why do we write this note?} It is erroneous to pretend to extract physical information from the experimental light curves (time series) of astrophysical systems by means of linear stochastic differential equations (LSDE). In general, the time evolution of these systems is governed by a set of nonlinear differential equations. Hence, the LSDEs are not suitable to model their dynamics. In spite of this, recently the LSDEs have been proposed as tools for the analysis of AGN light curves.
Their use in this context seems to be dictated by their simplicity rather than by a real physical argument. We stress in this note that the correct approach to the analysis of signals coming from systems with nonlinear dynamics is to tackle the problem using methodologies in well defined physical contexts.}

\section{SDEs in astronomy}

The main aims in the analysis of a time series are forecasting and modeling. Although the latter could imply the former, the reverse is not true. For example, what matters a broker is a statistical-mathematical tool which permits him to forecast the future value of a given stock index. However, even in the case of correct predictions, such tool (e.g. a neural network) could have no relationship with the underlying dynamics. The correct forecast of a time series does not necessarily mean to have understood its true dynamics. A trivial example is represented by an object moving on the X-Y plane along a circular orbit of radius $r$ around a central point with constant angular velocity. It is easy to realize that two observers, one located on the Z-axis and the other on the X-Y plane at a distance greater than $r$ from the origin, can exactly forecast the time evolution of this system. However, while the first observer can correctly realize that the observed dynamics is due a uniform circular motion, the second observer can only conclude that the system evolves according to a harmonic motion. The point is that a time series provides information only on a projection of the dynamics of the system under investigation. Lacking of any additional information, the reconstruction of the true dynamics from one projection is not possible. 

An example of the differences between forecasting and modeling in astronomy is represented by the W\"olfer sunspot number. A huge literature exists on the forecasting of this number by means of statistical models such as AR, ARMA or more sophisticated tools as the neural networks  (try a web search using``sunspot forecast'' or  ``sunspot prediction''). Here,  independently of the understanding of the involved physical processes, a reliable forecast is vital for the telecommunications since the W\"olfer number is linked to 
the solar activity. Of course, predictions based on a physical dynamical model, as it happens with the weather forecast, should be preferable. But this is a much more complex issue. Physical models for the dynamics of the W\"olfer sunspot number are available in literature, but they are not yet reliable for predictions (e.g. see \cite{all10, cam17} and reference therein). Prediction is not the main scope of such models, but rather to get insights on the physical processes driving the time evolution of the investigated system.

In astronomy and astrophysics the main aim of the analysis of a time series is to understand why a given system behaves as it is observed. For this reason the use of statistical-mathematical tools developed for the forecasting makes no sense \cite{vio05}.
The main limit in modeling the time series of astrophysical systems is that many of the involved physical processes are unknown. A possible way out is to assume that the dynamics of the system is driven by a small number of dominant physical 
processes whereas the ensemble of the unknown processes can be considered to constitute a stochastic perturbation.  The rationale of such an approach is that the unknown processes usually are due to the interaction of the physical system of interest with its surroundings and/or the action of complex processes that cannot be directly included in the model (e.g. gas turbulence). In general, such processes are characterized by a huge number of degrees of freedom and therefore they can be assumed to have a stochastic nature. In practice, this means study of the time evolution of a given physical system in the context of the so called stochastic dynamics, i.e., through the modeling of the observed time series by means of stochastic differential equations (SDE). This approach is largely followed in many branches of science and engineering (e.g. \cite{dua2005}). Surprisingly this does not happen in astronomy where often statistical-mathematical models are still adopted without specifying the physical reason. This is the case of two recent papers \cite{kas17, mor18} where the authors propose to use a LSDE  called CARMA, which represents the continuous version of the classical discrete stochastic linear ARMA model, to extract information from the variability of AGNs. 
The point is that AGNs are certainly (probably highly) nonlinear systems. Hence, it is not clear the utility of LSDE  to describe the time evolution of a nonlinear system. Actually, in appendix A of \cite{kas17} authors try to justify the use of the CARMA model as a consequence of the linearization of the true nonlinear dynamical equations due the small amplitude of the perturbative processes. However, this is a fact that should be proved and not a priori assumed.

As explained in \cite{vio05}, an effective use of the SDEs for the analysis of experimental signals requires two distinct operations: a) fit of specific dynamical models to the time series; b) validation of the results through the generation of synthetic signals to compare with the experimental ones. These operations require the capability to estimate the parameters in SDE from discrete observations and the numerical solution of this kind of equations. Although 
the numerical integration of the SDEs is a well developed topic since many years (e.g. see \cite{klo94, hig01, mil04}), such tasks could appear problematic for non-experts of the field. Actually, nowadays
free software is available not only for the numerical integration of the SDEs but also for their fit to experimental discrete signals (e.g. see the R packages Yuima and CTSMR).

\section{Final remarks}

The time series usually available in astronomy are able to characterize only a subset of the system of equations that describe the dynamics of the physical system under study. For this reason, although in principle it is always possible to find a
statistical model able to reproduce the experimental data, without any a priori physical model there are not many possibilities to obtain a reliable reconstruction of the physical scenario investigated. In general, this means that an approach to the analysis
of time series exclusively based on the experimental data will provide inconclusive results, and that the practice to search for more and more sophisticated statistical techniques is not productive. In many situations, the only possibility for physical
insights is to carry out the analysis in a well-defined physical context.

\end{document}